\documentclass[aps,pre,reprint,showpacs]{revtex4-1}

\usepackage{verbatim}
\usepackage{graphicx}
\usepackage{amsmath}
\usepackage{amssymb}
\usepackage[usenames, dvipsnames]{color}
\usepackage[normalem]{ulem}

\begin{document}

\preprint{}

\title{Optimal protocols for slowly-driven quantum processes}

\author{Patrick R. Zulkowski}
\email[]{pzulkowski@berkeley.edu}
\affiliation{Department of Physics, University of California, Berkeley, California 94720, USA}
\affiliation{Department of Mathematics, Berkeley City College, Berkeley, California 94704, USA}
\affiliation{Redwood Center for Theoretical Neuroscience, University of California, Berkeley, California 94720, USA}

\author{Michael R. DeWeese}
\email[]{deweese@berkeley.edu}
\affiliation{Department of Physics, University of California, Berkeley, California 94720, USA}
\affiliation{Redwood Center for Theoretical Neuroscience, University of California, Berkeley, California 94720, USA}
\affiliation{Helen Wills Neuroscience Institute, University of California, Berkeley, California 94720, USA}

\begin{abstract}
The design of efficient quantum information processing will rely
on optimal nonequilibrium transitions of driven quantum systems. Building on a recently-developed geometric framework for computing optimal protocols for classical systems driven in finite-time, we construct a general framework for optimizing the average information entropy for driven quantum systems. Geodesics on the parameter manifold endowed with a positive semi-definite metric correspond to protocols that minimize
the average information entropy production in finite-time. We use this framework to explicitly compute the optimal entropy production for a simple two-state quantum system coupled to a heat bath of bosonic oscillators, which has applications to quantum annealing.   \end{abstract}


\date{\today}

\maketitle

\section{Introduction}

Though a unifying set of principles describing all known nonequilibrium phenomena remains undiscovered, many recent developments have illuminated the thermodynamic behavior of small-scale systems. For instance, fluctuation theorems valid far from equilibrium have been developed in the classical setting~\cite{Evans1993,Evans1994,Gallavotti1995a,Crooks1999,Hatano2001,Jarzynski1997,Liphardt2002,Seifert2005b,Sagawa2010,Wang2002,Carberry2004,Garnier2005,Toyabe2010} as well as in the quantum regime~\cite{Lutz2011,Liu2014,Leggio2013,Chetrite2012,Albash2013,Crooks2008,Campisi2011,Campisi2009,Talkner2009}.

An area of nonequilibrium thermodynamics of particular interest concerns the operation of small-scale information processing systems. The interplay between information as a physical quantity and thermodynamics has a rich history~\cite{Szilard1929,Landauer1961,Bennett1982}. 

The physics of information processing is of particular relevance considering the rapid development of information technology and the inevitable approach to computational limits imposed by physical law~\cite{Frank2002,Lambson_2011}. Optimization schemes for small-scale information processing occurring in finite time will be needed to develop technology capable of approaching those limits~\cite{Andresen2011,Chen2004}.

Current research has uncovered techniques to optimize thermodynamic quantities arising in small-scale systems designed to store and erase classical information~\cite{Diana2013,Esposito2010,Aurell2012,Zulkowski:2014gz}, including the derivation of a refined second law~\cite{Aurell2012}. This research couples with the progress made on the general problem of predicting optimal protocols to drive classical systems between stationary states with minimal dissipation~\cite{Sivak:2012gr,Zulkowski:2012if,Zulkowski:2013fu,Zulkowski:2014gz, Shenfeld2009,Brody2009,Seifert2008,Seifert2007,Aurell2011}. 

In parallel with classical developments, a greater understanding of optimal processes in the nonequilibrium quantum regime and the efficiency of quantum engines has been achieved~\cite{Abah2014,Mehta2013,Deffner2014,Stefanatos2014,Schmiedl2009}. The success of a recently-proposed linear-response framework for optimal driving of microscopic classical systems~\cite{Sivak:2012gr} calls out for an extension to quantum systems. 

In the geometric formulation of~\cite{Sivak:2012gr}, a generalized inverse diffusion tensor induces a Riemannian manifold structure on the space of parameters, and optimal protocols trace out geodesics of this inverse diffusion tensor. This geometric framework is subsequently developed and exploited in~\cite{Zulkowski:2012if,Zulkowski:2013fu,Zulkowski:2014gz}.

In this paper, we extend this work to provide a geometric framework for computing control protocols optimizing the average information entropy production~\cite{Lutz2011}. The production of entropy is intimately related to the overall performance of thermodynamic devices by in some sense quantifying irreversibility and providing a bound on the availability of useful work. Entropy production also provides a useful tool in the analysis of nonequilibrium effects. 

By twice measuring the density operator of a system interacting with a thermal reservoir at the beginning and end of the protocol, the average information entropy production may be defined and shown to satisfy a fluctuation theorem~\cite{Lutz2011}. While there is still debate about the best way to define thermodynamic quantities along quantum trajectories, this approach provides an avenue for experimental observation~\cite{Lutz2011}. Fortunately, this formalism holds for open quantum systems driven arbitrarily far from equilibrium. 

We begin by constructing a general positive semi-definite tensor on the space of control parameters for the quantum system interacting with the thermal reservoir. We assume the dynamics of the system are described by a master equation of Lindblad form, arising from an adiabatic, rotating-wave approximation in a sense we make concrete below~\cite{Albash2012}. 

With the general tensor in hand, we compute optimal protocols for a simple two-state system coupled to a thermal bath of bosonic oscillators. The system Hamiltonian may be interpreted as describing a spin-$\frac{1}{2} $ particle coupled to a time-dependent magnetic field with components in the $ y $ and $ z $-directions. This system has applications in quantum annealing~\cite{Albash2012, Kadowaki1998}. For this simple system, we demonstrate the existence of null directions of the metric tensor, which correspond to directions in parameter space in which there is no overall change in information entropy. We derive an approximate expression for the optimal overall entropy production.

\section{The quantum tensor}

Our model consists of two distinct components: the system under our control and a large collection of quantum degrees of freedom acting as the thermal reservoir. Together, the system and the bath degrees of freedom evolve unitarily according to the von Neumann equation $ \partial_{t} \rho_{tot} = - \frac{i}{\hbar} [ \hat{H}_{tot} ,\rho_{tot} ]$, where 
\begin{equation} \hat{H}_{tot} = \hat{H}_{sys}(t) + \hat{H}_{B} + g \sum_{\alpha} A_{\alpha } \otimes B_{\alpha}. \end{equation}
The operator $ \hat{H}_{sys}(t) $ is the system Hamiltonian acting in the Hilbert space of system states while $ \hat{H}_{B} $ is the Hamiltonian for the bath degrees of freedom. The interaction term $ g \sum_{\alpha} A_{\alpha } \otimes B_{\alpha} $ consists of a (weak) coupling $ g $ and Hermitian operators $ A_{\alpha} $ and $ B_{\alpha} $ operating on the system and bath Hilbert spaces respectively. 

We are only interested in the time evolution of the density operator of the system, denoted $ \rho_{t} $, which can be obtained from $ \rho_{tot} $ by tracing over the bath degrees of freedom. We follow the construction of~\cite{Albash2012}, which obtains a quantum Markovian master equation governing $ \rho_{t} $ via an adiabatic, rotating-wave approximation. Specifically, the derivation of Eq.~\eqref{eq:Lindblad} in~\cite{Albash2012} utilizes the so-called ``standard adiabatic approximation"
\begin{equation}\label{eq:adiabaticapprx}
\frac{h}{\triangle^2 \tau} \ll 1 
\end{equation}
where $\tau$ denotes the total evolution time, $ \triangle \equiv \displaystyle \min_{\substack{t \in [0,\tau]}} \left \{ \epsilon_{1}(t) - \epsilon_{0}(t) \right \}$ is defined to be the minimum ground state energy gap of $ H_{sys} $, and 
\begin{equation} h \equiv \displaystyle \max_{\substack{s \in [0, 1]; a , b}} |\langle \epsilon_a(s) | \partial_{s} H_{sys}(s) | \epsilon_{b}(s) \rangle. \end{equation} 
Here, $ s = t/\tau $ is a dimensionless measure of time. Though it is possible to compute higher-order terms in $ 1/\tau $~\cite{Albash2012}, we will assume that Eq.~\eqref{eq:Lindblad} adequately approximates the time-evolution of the quantum system over finite, but sufficiently long time scales.

Assuming a weak coupling $ g $ between the system and bath degrees of freedom, we have a master equation in Lindblad form:
\begin{align}\label{eq:Lindblad} \partial_{t} \rho_{t} 
=
 - \frac{i}{\hbar} [ \hat{H}_{t} , \rho_{t} ] \nonumber + g^2 \mathcal{L}(\rho_{t}),\end{align}
where 
\begin{align}
\mathcal{L}(\rho_{t}) = \sum_{\alpha \beta } \sum_{\omega} \gamma_{\alpha \beta} (\omega) \bigg( & L_{\omega,\beta}(t) \rho_{t} L_{\omega,\alpha}^{\dag}(t) - \nonumber \\ & \frac{1}{2} \{ L_{\omega,\alpha}^{\dag}(t) L_{\omega,\beta}(t), \rho_{t} \} \bigg).
\end{align}
Here, $ \hat{H}_{t} \equiv \hat{H}_{sys}(t) + g^2 \hat{H}_{LS}(t) $ consists of the system Hamiltonian $ \hat{H}_{sys}(t) \equiv  \hat{H}_{sys}(\boldsymbol \lambda (t)) $ and the Lamb shift Hamiltonian, which arises through the coupling of the system with the thermal reservoir. 

Assume that the system Hamiltonian has time-dependent eigenvalues $ \epsilon_{a}(t) $ with time-dependent eigenkets $ |\epsilon_{a}(t) \rangle $. Then the operators $ L_{\omega,\alpha}(t) $ are defined by
\begin{equation} L_{\omega,\alpha}(t) \equiv \sum_{\epsilon_{b}(t)-\epsilon_{a}(t) = \omega } L_{ab,\alpha}(t) \end{equation}
with
\begin{equation} L_{ab,\alpha}(t) \equiv \langle \epsilon_{a}(t) | A_{\alpha} | \epsilon_{b}(t) \rangle |\epsilon_{a}(t) \rangle \langle \epsilon_{b}(t) | .\end{equation}
Furthermore, 
\begin{equation}\label{eq:LambShift} \hat{H}_{LS} = \sum_{\alpha \beta} \sum_{\omega} L_{\omega,\alpha}^{\dag}(t) L_{\omega, \beta}(t) S_{\alpha \beta}(\omega) \end{equation}
where both $ \gamma_{\alpha \beta}(\omega) $ and $ S_{\alpha \beta }(\omega) $ are Hermitian and can be computed from the spectral-density matrix~\cite{Albash2012}. To ease notation we will suppress the time dependence of the Lindblad operators $ L_{\omega,\alpha} $.

It is crucial to note that the time dependence of the terms defining $ \mathcal{L} $ arises only through the time dependence of the spectrum and eigenkets of the system Hamiltonian. Therefore, the time dependence of the Lindblad operator $ \mathcal{L} $ stems from the control parameters $ \boldsymbol \lambda (t)$. If time appeared explicitly in the terms defining $ \mathcal{L} $, we could not interpret the approximation developed in this section as giving rise to a semi-definite metric on the space of control parameters. 

In what follows, we denote  the control parameter protocol by $ \boldsymbol \Lambda $. We assume the protocol to be sufficiently smooth to be twice-differentiable. The framework in the classical setting is versatile and can handle situations in which jump discontinuities are present~\cite{Zulkowski:2014gz}. Jump discontinuities at the end points of the protocol commonly arise in optimal finite-time driving processes of classical systems~\cite{Seifert2007,Seifert2008,Aurell2011,SeifertReview_2012,Esposito2010,Aurell2012}. For simplicity we only admit twice-differentiable protocols though one could in principle extend this approach to piecewise-continuous $ \boldsymbol \Lambda $.

By definition~\cite{Lutz2011}, the average information entropy is given by
\begin{equation} \langle \Sigma_{I} \rangle_{\boldsymbol \Lambda} = \int_{0}^{\tau} dt~\mbox{Tr} \big \{ \partial_{t} \rho \big[ - \ln \rho + \ln \rho_{t}^{eq} \big] \big \},\end{equation}
where 
\begin{equation} \rho_{t}^{eq} = \frac{e^{- \beta \hat{H}_{sys}(\boldsymbol \lambda (t))}}{ \mbox{Tr} \{ e^{- \beta \hat{H}_{sys}(\boldsymbol \lambda (t))}\} } \end{equation}
is the equilibrium distribution defined by the instantaneous control parameters $ \boldsymbol \lambda(t) $ and $ \beta $ is related to the thermal bath temperature by $ \beta = \left( k_{B} T \right)^{-1} $ where $ k_{B} $ is Boltzmann's constant. 
Using Eq.~\eqref{eq:Lindblad}, we see
\begin{align}\label{eq:averageentropy} \langle \Sigma_{I} \rangle_{\boldsymbol \Lambda}  &= \int_{0}^{\tau} dt~ \mbox{Tr} \bigg \{ \left ( - i [ \hat{H}_{t} , \rho_{t}]+ g^2 \mathcal{L}(\rho_{t}) \right ) \times \nonumber \\ & \left( - \ln \rho_{t} + \ln \rho_{t}^{eq} \right) \bigg \} \nonumber \\
&= g^2 \int_{0}^{\tau} dt~ \mbox{Tr} \bigg \{\mathcal{L}(\rho_{t}) \left( - \ln \rho_{t} + \ln \rho_{t}^{eq} \right) \bigg \},
\end{align}
where $\tau$ represents the duration of the protocol.
This equation follows from the cyclic property of the trace:

The trace term involving $ - \ln \rho_{t} $ and the commutator vanishes because we may permute $ - \ln \rho_{t} $ into the commutator with $ \rho_{t} $. As for the $ \rho_{t}^{eq} $ term, it too can be permuted into the commutator and $ [\hat{H}_{sys}, \rho_{t}^{eq} ] = 0 $, naturally. A little more work goes into showing that $ [\hat{H}_{LS}, \rho_{t}^{eq} ] = 0 $. Since $ \rho_{t}^{eq} = \frac{e^{-\beta \hat{H}_{sys}(\boldsymbol \lambda(t))}}{\mbox{Tr} \left \{ e^{-\beta \hat{H}_{sys}(\boldsymbol \lambda(t))}\right \} }$ by definition, it is sufficient to show that $  [\hat{H}_{LS}, \hat{H}_{sys}(\boldsymbol \lambda(t)) ] = 0 $.

From Eq.~\eqref{eq:LambShift}, we need only establish that $ [ L_{\omega,\alpha}^{\dag} L_{\omega, \beta}  , \hat{H}_{sys}(\boldsymbol \lambda(t))  ] = 0 $. First, note that
\begin{align} [ L_{\omega,\beta} , \hat{H}_{sys} ] & = \left [ \sum_{\epsilon_b - \epsilon_a = \omega} \langle \epsilon_a | \hat{A}_{\beta} | \epsilon_{b} \rangle |\epsilon_a \rangle \langle \epsilon_b |, \hat{H}_{sys}\right ] \nonumber \\ & =   \sum_{\epsilon_b - \epsilon_a = \omega} \langle \epsilon_a | \hat{A}_{\beta} | \epsilon_{b}  \rangle \left [ |\epsilon_a \rangle \langle \epsilon_b |, \hat{H}_{sys} \right ] \nonumber \\ & = \sum_{\epsilon_b - \epsilon_a = \omega} \langle \epsilon_a | \hat{A}_{\beta} | \epsilon_{b}  \rangle \left(\epsilon_{b}-\epsilon_{a} \right) |\epsilon_a \rangle \langle \epsilon_b | \nonumber \\ & = \omega L_{\omega,\beta} \end{align}
where we have suppressed the time dependence for brevity. This further implies that
\begin{equation} [ L_{\omega,\alpha}^{\dag} , \hat{H}_{sys} ] = - \left( [ L_{\omega,\alpha} , \hat{H}_{sys} ]  \right)^{\dag} = - \omega L_{\omega,\alpha}^{\dag}  \end{equation}
and so 
\begin{align} [ L_{\omega,\alpha}^{\dag} L_{\omega, \beta}  , \hat{H}_{sys}(\boldsymbol \lambda(t))  ]  & = L_{\omega,\alpha}^{\dag}   [ L_{\omega, \beta}  , \hat{H}_{sys}(\boldsymbol \lambda(t))  ] \nonumber \\ & + [ L_{\omega,\alpha}^{\dag} , \hat{H}_{sys}(\boldsymbol \lambda(t))  ] L_{\omega, \beta} \nonumber \\ & =  \omega L_{\omega,\alpha}^{\dag}    L_{\omega, \beta} - \omega L_{\omega,\alpha}^{\dag}    L_{\omega, \beta} \nonumber \\ & = 0   \end{align} 
establishing our claim.

We see from Eq.~\eqref{eq:averageentropy} that the average information entropy is proportional to $ g^2 $. This seems reasonable since if $ g = 0 $, then there would be no coupling between the system and the bath. The system would then evolve in time unitarily and the average entropy would vanish.

This observation allows us to drastically simplify the mathematics since the expressions inside the integral need only be kept to $ 0 $-th order in $ g^2 $. In other words, we may compute the evolution of $ \rho_{t} $ using only the von Neumann equation 
\begin{equation}\label{eq:vonNeumann} \partial_{t} \rho_{t} = - \frac{i}{\hbar} [ \hat{H}_{sys}(t), \rho_{t} ]. \end{equation}

We wish to approximate Eq~\eqref{eq:averageentropy} when the protocol duration $ \tau $ is large in the sense of Eq.~\eqref{eq:adiabaticapprx}. To achieve this end we utilize the so-called derivative truncation method~\cite{Zulkowski:2012if,Zulkowski:2013fu,Zulkowski:2014gz}, which assumes a specific form for the density operator in terms of the equilibrium system density operator and the first order derivative of the protocol $ \boldsymbol{\Lambda}$:
\begin{equation}\label{eq:derivapprox} \rho_{t} \approx \rho_{t}^{eq} + \delta \rho_{\lambda^{\alpha}} \frac{d \lambda^{\alpha}}{dt}. \end{equation}
The Einstein summation convention is assumed here and throughout for the index $ \alpha $ and the operator $ \delta \rho_{\lambda^{\alpha}} $ is Hermitian and traceless. 

Substituting Eq.~\eqref{eq:derivapprox} into the von Neumann equation Eq.~\eqref{eq:vonNeumann} and ignoring derivative terms of order higher than first, we obtain equations for the unknown operators  $ \delta \rho_{\lambda^{\alpha}} $ where $ \alpha $ indexes the finite set of control parameters.

These equations are most conveniently expressed in terms of the operator basis $ | \omega_{a}(t) \rangle \langle \omega_{b}(t) | $ where $ |\omega_{a}(t) \rangle $ is an eigenket of $ \rho_{t} $ with eigenvalue $ \omega_{a} $. The convenience of this choice arises in the time independence of the eigenvalues $ \omega_{a} $, which can be illustrated by the following argument:

As $ \rho_{t} $ evolves according to Eq.~\eqref{eq:vonNeumann}, we must have 
\begin{align} \partial_{t} \mbox{Tr} \left \{ \rho_{t}^{n} \right \}  = \mbox{Tr} \left \{ n \left( - \frac{i}{\hbar}  [ \hat{H}_{sys}(t), \rho_{t} ] \right) \rho_{t}^{n-1} \right \} =0\end{align}
for all positive integers $ n $. Since the coefficients of the characteristic polynomial of $ \rho_{t} $ can be expressed in terms of combinations of traces of powers of $ \rho_{t} $, it follows that the spectrum is time independent; i.e.; $ \omega_{a}(t) = \omega_{a}(0) \equiv \omega_{a} $. 

It follows immediately that 
\begin{equation} \partial_{t} \rho_{t} = \sum_{a} \omega_{a} \partial_{t} \big [ | \omega_{a}(t) \rangle \langle \omega_{a}(t) | \big ] \end{equation}
by the Spectral Theorem. Using the eigenket basis $ |\omega_{a}(t) \rangle $ affords us a simple expression for $ \partial_{t} \rho_{t} $: we need only compute the time-derivative of the projection operators since the eigenvalues $ \omega_{a} $ are time-independent. However, the (time-dependent) energy eigenkets are more convenient for practical applications and so we express all relevant quantities in terms of this basis. 

Using the derivative truncation approximation, we can deduce the approximate eigenkets of $ \rho_{t} $ in terms of the energy eigenkets:
\begin{align}\label{eq:eigenkets} | \omega_{a} \rangle & \approx |\epsilon_a \rangle + \sum_{b \neq a } \left( \frac{Z_{t}}{e^{-\beta \epsilon_{a} } - e^{-\beta \epsilon_b}} \right) \langle \epsilon_b | \delta \rho_{\lambda^{\alpha}} | \epsilon_a \rangle | \epsilon_b \rangle \frac{d \lambda^{\alpha}}{dt} \nonumber \\& \equiv |\epsilon_a \rangle + \left( \delta |\omega_a \rangle \right)_{\lambda^{\alpha}} \frac{d \lambda^{\alpha}}{dt} . \end{align}
Furthermore, $ \omega_{a} \approx \frac{e^{-\beta \epsilon_a}}{Z} $. This follows immediate from the approximation Eq.~\eqref{eq:derivapprox} and the definition of eigenkets. 

The construction of the approximate eigenkets and eigenvalues forces the diagonal entries of $ \delta \rho_{\lambda^{\alpha}} $ in the energy eigenket basis to vanish. This is consistent with the requirement of positivity as both $ \rho_{t} $ and $ \rho_{t}^{eq} $ are positive in Eq.~\eqref{eq:derivapprox}, but the term involving $ \delta \rho_{\lambda^{\alpha}} $ is sensitive to the rate of change of $ \lambda^{\alpha} $ which could be negative.

Using Eqs.~\eqref{eq:derivapprox} and~\eqref{eq:eigenkets} , we see that
\begin{equation} \partial_{t} \rho_{t} \approx \sum_{a}  \frac{e^{-\beta \epsilon_a}}{Z_{t}} \frac{\partial}{\partial \lambda^{\alpha}} \big[ | \epsilon_{a} \rangle \langle \epsilon_{a} | \big] \frac{d \lambda^{\alpha}}{dt} \end{equation}
and so we obtain the expansion of $ \delta \rho_{\lambda^{\alpha}} $ in the energy eigenket operator basis: 
\begin{align} \delta \rho_{\lambda^{\alpha}} & = \sum_{ a \neq c } \sum_{b}  \left( \frac{\hbar i e^{-\beta \epsilon_b}}{\epsilon_{ac}Z_{t}} \mbox{Tr} \left \{ | \epsilon_c \rangle \langle \epsilon_a | \partial_{\lambda^{\alpha}} \left[| \epsilon_b \rangle \langle \epsilon_b |  \right] \right \} \right)  | \epsilon_a \rangle \langle \epsilon_c | \nonumber \\ & = \sum_{a \neq c} \frac{\hbar i}{\epsilon_{ac}} \left( \partial_{\lambda^{\alpha}} \rho^{eq}_{t} \right)_{ac} | \epsilon_a \rangle \langle \epsilon_c |  \end{align} 
with $  \epsilon_{ac} \equiv \epsilon_{a}-\epsilon_{c} $ and $ \left( \partial_{\lambda^{\alpha}} \rho^{eq} \right)_{ac} \equiv \text{Tr} \left \{ | \epsilon_c \rangle \langle \epsilon_a | \partial_{\lambda^{\alpha}} \rho^{eq}_{t} \right \} $. 

Furthermore, from Eq.~\eqref{eq:eigenkets} we have
\begin{equation}\label{eq:eigenketssimp}
\left( \delta |\omega_a \rangle \right)_{\lambda^{\alpha}} = \sum_{b \neq a} \frac{Z_{t}}{e^{-\beta \epsilon_a}-e^{-\beta \epsilon_{b}}} \frac{\hbar i}{\epsilon_{ba}} \left( \partial_{\lambda^{\alpha}} \rho^{eq}_{t} \right)_{ba} | \epsilon_b \rangle. \end{equation}

By constructing an explicit expression for $ \delta \rho_{\lambda^{\alpha}}$ we may now construct a quadratic functional approximating $ \langle \Sigma_{I} \rangle_{\boldsymbol \Lambda} $:
\begin{align} \langle \Sigma_{I} \rangle_{\boldsymbol \Lambda} & = g^2 \int_{0}^{\tau} dt~ \mbox{Tr} \left \{ \mathcal{L}(\rho_{t}) \left( - \ln \rho_{t} + \ln \rho_{t}^{eq} \right) \right \} \nonumber \\ & \approx g^2 \int_{0}^{\tau} dt~\frac{d \lambda^{\beta}}{dt} \mbox{Tr} \left \{ \mathcal{L}(\delta \rho_{\lambda^{\beta}}) \left( - \ln \rho_{t}  + \ln \rho_{t}^{eq} \right) \right \} \end{align}
which follows from $ \mathcal{L}(\rho_{t}^{eq}) = 0 $. 

We compute the trace in the $ | \omega_{a} \rangle $ basis in which $ - \ln \rho_{t} $ is diagonal. Since $ \mbox{Tr} \{ \mathcal{L} \left( \cdot \right) \}  = 0 $, we have
\begin{align} \frac{d \lambda^{\beta}}{dt} & \mbox{Tr} \left \{ \mathcal{L} \left( \delta \rho_{\lambda^{\beta}} \right) \left( - \ln \rho_{t} \right) \right \}  \nonumber \\ & =  \frac{d \lambda^{\beta}}{dt} \mbox{Tr} \left \{ \mathcal{L} \left( \delta \rho_{\lambda^{\beta}} \right) \left( - \ln \rho_{t} + \ln Z_{t} \right) \right \} \nonumber\\ & \approx \frac{d\lambda^{\beta} }{dt}  \mbox{Tr} \left \{ \mathcal{L} \left( \delta \rho_{\lambda^{\beta}} \right) \left( - \ln \rho_{t}^{eq} \right) \right \} \nonumber \\ & + \sum_{a} 2 \beta \epsilon_{a}~\Re \left \{ \langle \epsilon_a | \mathcal{L}(\delta \rho_{\lambda^{\beta}}) \delta \left( |\omega_{a} \rangle \right)_{\lambda^{\alpha}} \right \} \frac{d \lambda^{\alpha}}{dt} \frac{d \lambda^{\beta}}{dt}. \end{align}

From this follows a useful expression for the entropy production
\begin{align}\label{eq:tensor} \langle \Sigma_{I} \rangle_{\boldsymbol \Lambda}   \approx g^2 \int_{0}^{\tau} & dt~\frac{d \lambda^{\alpha}}{dt} \frac{d \lambda^{\beta}}{dt}  \times \nonumber \\ & \sum_{a} \beta \epsilon_{a}~\Re  \{ \langle \epsilon_a | \mathcal{L}(\delta \rho_{\lambda^{\beta}}) \delta \left( |\omega_{a} \rangle \right)_{\lambda^{\alpha}} \nonumber \\ & +\langle \epsilon_a | \mathcal{L}(\delta \rho_{\lambda^{\alpha}}) \delta \left( |\omega_{a} \rangle \right)_{\lambda^{\beta}}  \}   \end{align} 
after symmetrization in the $ \alpha $ and $ \beta $ indices. We may write this expression as
\begin{equation}\label{eq:tensorsimp}
\langle \Sigma_{I} \rangle_{\boldsymbol \Lambda}   \approx g^2 \int_{0}^{\tau} dt~\sum_{jklm} \frac{d \lambda^{\alpha}}{dt} \frac{d \lambda^{\beta}}{dt}\mathcal{A}_{jklm} \left( \partial_{\lambda^{\alpha}} \rho^{eq}_{t} \right)_{jk} \left( \partial_{\lambda^{\beta}} \rho^{eq}_{t} \right)_{lm} 
\end{equation}
where the components of the matrix $ \mathcal{A} $ depend on $ \mathcal{L} $ and the energy eigenvalues and eigenkets. We include the explicit components of $ \mathcal{A} $ in the appendix.

Eq.~\eqref{eq:tensor} approximates the average information entropy produced during a finite-time driving protocol of a quantum system weakly coupled to a large thermal bath using only quantities directly calculable from the time-dependent system Hamiltonian and the Lindblad operators.

We will now use Eq.~\eqref{eq:tensor} to explicitly compute the average information entropy produced by driving a simple two-state model quantum system.

\section{Two-state model system}

We apply Eq.~\eqref{eq:tensor} to a simple two-state system with system Hamiltonian
\begin{equation} \hat{H}_{sys} = -\hbar \omega_{y}(t) \sigma_{y} + \hbar \omega_{z}(t) \sigma_{z} \end{equation}
where $ \sigma_{y} $ and $ \sigma_{z} $ are the Pauli spin matrices. Optimization of such a system could potentially be useful in applications such as quantum annealing~\cite{Albash2012, Kadowaki1998}.

We further assume that the system is coupled to a thermal bath of bosonic oscillators so that the full Hamiltonian is
\begin{equation}  \hat{H}_{tot} = \hat{H}_{sys}(t) + g~\sigma_{z} \otimes \hat{B} + \hat{H}_{B}\end{equation}
where $ \hat{H}_{B} = \sum_{k} \omega_{k} b_{k}^{\dag} b_{k} $, $ \hat{B} = \sum_{k} \left( b_{k}^{\dag} + b_{k} \right) $ and $ g $ is a small coupling constant.

It is convenient to work in an eigenbasis of $ \sigma_{z} $ represented by the column vectors 
\begin{equation} | \uparrow \rangle = \left( \begin{array}{c}
1 \\
0  \end{array} \right) ,  | \downarrow \rangle = \left( \begin{array}{c}
0 \\
1  \end{array} \right). \end{equation}
In this basis,
\begin{equation} \hat{H}_{sys} = \hbar\left( \begin{array}{cc}
\omega_{z} & i \omega_{y}  \\
- i \omega_{y} & -\omega_{z}  \end{array} \right). \end{equation}

It is most convenient to express the control parameters in polar form: $ \beta \hbar \omega_{z} \equiv r \cos(\theta) , \beta \hbar \omega_{y} \equiv r \sin(\theta)$. 
In this form, it is not hard to show that the eigenvalues of $ \hat{H}_{sys} $ are $ \mp r $ with eigenvectors

\begin{equation} | -r \rangle = \left( \begin{array}{c}
- i \sin \left( \theta/2 \right) \\
\cos \left (\theta/2 \right)  \end{array} \right) ,  | r \rangle = \left( \begin{array}{c}
i \cos \left( \theta/2 \right) \\
\sin \left (\theta/2 \right)  \end{array} \right). \end{equation}

For simplicity, we assume $ r > 0 $ and $ 0 < \theta < \frac{\pi}{2} $. A straightforward calculation demonstrates that
\begin{equation} L_{0} = \left( \begin{array}{cc} \cos^2 (\theta) & i \cos(\theta) \sin(\theta) \\ - i \cos(\theta) \sin(\theta) & -\cos^2(\theta) \end{array} \right)\end{equation}
\begin{equation} L_{-2r} = \left( \begin{array}{cc}  \frac{1}{2} \sin^2(\theta)  & -i \cos^2(\theta/2) \sin(\theta) \\ - i \sin^2(\theta/2) \sin(\theta) & -\frac{1}{2} \sin^2(\theta) \end{array} \right) \end{equation}
\begin{equation} L_{2r} = \left( \begin{array}{cc} \frac{1}{2} \sin^2(\theta)  & i \sin^2(\theta/2) \sin(\theta) \\  i \cos^2(\theta/2) \sin(\theta) & -\frac{1}{2} \sin^2(\theta) \end{array} \right)\end{equation}
are the operators defining $ \mathcal{L} $ for this model system. 

Choosing the bath oscillator frequencies $ \omega_{k} $ so that $ \gamma(0) \equiv \gamma_{0}  $ and $ \gamma(\pm 2 r ) = \gamma_{1} e^{\pm r }  $ for $ \gamma_{0} , \gamma_{1} > 0 $ and setting $ \hbar =1 $, we have from Eq.~\eqref{eq:tensor}
\begin{align} \langle \Sigma_{I} \rangle_{\boldsymbol \Lambda} & \approx g^2  \int_{0}^{\tau} \Bigg[ \gamma_{0} \frac{\left(x_{0}- \frac{1}{2} \right) \ln \left(\frac{x_{0}}{1-x_{0}} \right)}{2 r^2 } \cos^2(\theta) \nonumber \\ & + \gamma_{1} \frac{ \left( e^{r}-e^{-r} \right)\ln \left(\frac{x_{0}}{1-x_{0}} \right) }{16 r^2 }\sin^2 \left( \theta \right)\Bigg]  \left( \frac{d \theta}{dt} \right)^2 dt \nonumber \\ 
& \equiv \int_{0}^{\tau} \sigma^2[\theta(t)]  \left( \frac{d \theta}{dt} \right)^2 dt \end{align}
where $  x_{0} = \frac{1}{1+e^{-2r} }  $. 

While this functional is non-negative as expected from the fluctuation theorem of~\cite{Lutz2011}, we see that it vanishes if $ \theta $ is held constant. If $ r $ is allowed to vary while $ \theta $ is fixed, no information entropy is generated on average. In terms of Riemannian geometry, this means that the metric tensor possesses null directions.

We apply the Euler-Lagrange equation to obtain the optimal entropy production:

\begin{equation} \frac{\partial}{\partial \theta} \big \{ \sigma^2[\theta] \dot{\theta}^2 \big \}- \frac{d}{dt} \left( \frac{\partial}{\partial \dot{\theta} } \big \{ \sigma^2[\theta]\dot{\theta}^2 \big \} \right) = 0   \end{equation} 
implies
\begin{equation} \ddot{\theta}+ \frac{ \sigma'[\theta]}{\sigma[\theta]} \dot{\theta}^2 = 0.    \end{equation} 
Upon integration, we find
\begin{equation} \dot{\theta} = \frac{ C}{\sigma[\theta]} \  , \ C = \frac{1}{\tau} \int_{\theta_{0}}^{\theta_{f}} \sigma[\theta] d \theta.  \end{equation}
Therefore, the optimal average entropy is given by
\begin{equation} \langle \Sigma_{I} \rangle_{\boldsymbol \Lambda_{\text{opt}}} \approx \frac{1}{\tau} \left( \int_{\theta_{0}}^{\theta_{f}} \sigma[\theta] d\theta \right)^2. \end{equation} 
If we choose $ \theta_{0} = 0 \ , \ \theta_{f} = \pi/2 $, then 
\begin{align}\label{eq:optentropy} \langle \Sigma_{I} \rangle_{\boldsymbol \Lambda_{\text{opt}}} \approx \frac{g}{\tau} & \sqrt{  \frac{\gamma_{0} \left(x_{0}- \frac{1}{2} \right) \ln \left(\frac{x_{0}}{1-x_{0}} \right)}{2 r^2 } } \times \nonumber \\ & \text{EllipticE} \left(1- \frac{\gamma_{1} \left( e^{r}+e^{-r} \right)}{ 8\gamma_{0}} \right) . \end{align} 

\begin{figure}[ht!]
\begin{center}
\includegraphics{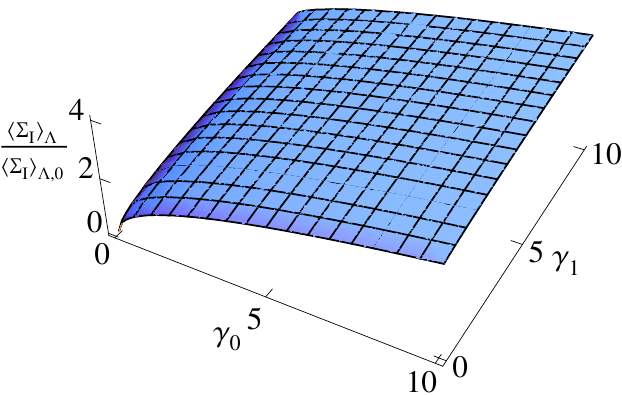}
\end{center}
\caption{Plot of $ \langle \Sigma_{I} \rangle_{\boldsymbol \Lambda_{\text{opt}}} / \langle \Sigma_{I} \rangle_{\boldsymbol \Lambda_{\text{opt}},0}$ where $ \langle \Sigma_{I} \rangle_{\boldsymbol \Lambda_{\text{opt}},0}$ is the optimal average entropy for $ \gamma_{0} = 1 \ , \ \gamma_{1} = 0$. }
\label{fig:optent}
\end{figure}

Eq.~\eqref{eq:optentropy} exhibits the expected $ 1/\tau $ behavior of the optimal average entropy production. Fig.~\ref{fig:optent} illustrates the dependence of $ \langle \Sigma_{I} \rangle_{\boldsymbol \Lambda_{\text{opt}}}  $ on the constants $ \gamma_{0} $ and $ \gamma_{1} $. We see that $\langle \Sigma_{I} \rangle_{\boldsymbol \Lambda_{\text{opt}}} $ has a $ \gamma_{1} = \text{constant} $ profile described by $ \sqrt{\gamma_{0}} $. The quantities $ \gamma_{0} $ and $ \gamma_{1} $ are related to the bosonic frequencies $ \omega_{k} $ of the thermal bath~\cite{Albash2012} and are thereby related to the noise of the quantum system. The bosonic frequencies consequently have a relatively simple influence on the overall average entropy production in the finite-time long duration limit via Eq.~\eqref{eq:optentropy}.

\section{Conclusion}

Using the formalism of~\cite{Sivak:2012gr,Zulkowski:2012if,Zulkowski:2013fu,Zulkowski:2014gz}, developed for classical systems, we were able to construct a general approximation of the average information entropy of a quantum system driven in finite-time in terms of a quadratic functional of velocities in parameter space. This functional can be interpreted as endowing parameter space with a semi-definite metric in which optimal protocols are equivalent to geodesics. For a simple two-state driven quantum system weakly coupled to a thermal bosonic bath, we were able to derive an approximate expression for the average information entropy. This expression has the characteristic $ 1/ \tau $ dependence with a coefficient compactly expressed in terms of quantities related to the bosonic bath oscillator frequencies.

Interestingly, in the simple two-state example studied here, 
the quadratic functional approximation for $ \langle \Sigma_{I} \rangle_{\boldsymbol \Lambda} $ we derived
possesses null directions when expressed in terms of metric geometry on the space of control parameters. 
In fact, for this model system, changing $ r $ results in a shift of energies. If $ r $ alone is changed, then it turns out that the system density matrix evolves exactly without a change in average information entropy.  Consequently, our solution contains a null direction, which makes the quantum tensor not positive definite but positive semi-definite. It seems likely that this behavior is generic, though a proof is lacking.

It is gratifying that this framework allowed us to obtain a general expression for the approximate entropy production in driven non-equilibrium quantum systems, as well as a closed-form solution for the minimum entropy production possible for a specific system with relevance to quantum annealing. We are encouraged by the success of this approach for this simple system and we hope that this program will lead to further insight into the optimization of quantum systems out of equilibrium.

\section{Acknowledgements.}
M.R.D. gratefully acknowledges
support from the McKnight Foundation and the Hellman Family
Faculty Fund. M.R.D.
and P.R.Z. were partly supported by the National Science
Foundation through Grant No. IIS-1219199. This material is
based upon work supported in part by the US Army Research
Laboratory and the US Army Research Office under Contract
No. W911NF-13-1-0390.

\section{Appendix}

We record here the components of $ \mathcal{A} $ appearing in Eq.~\eqref{eq:tensorsimp} for the convenience of the reader.

\begin{align}
&\mathcal{A}_{jklm} = \frac{1}{2} \sum_{b>a, d > c} \frac{ \hbar \beta Z_{t} }{\left( e^{-\beta \epsilon_{b}}-e^{-\beta \epsilon_{a}} \right) \epsilon_{dc}} \times \{ \nonumber \\ & - \left( \delta_{jd}~ \delta_{kc}~ \delta_{lb} ~\delta_{ma} + \delta_{ld}~ \delta_{mc} ~\delta_{jb} ~\delta_{ka} \right) \langle \epsilon_a | \mathcal{L}( |\epsilon_d \rangle \langle \epsilon_c |) | \epsilon_b \rangle \nonumber \\ & 
+\left( \delta_{jc}~ \delta_{kd}~ \delta_{lb} ~\delta_{ma} + \delta_{lc}~ \delta_{md} ~\delta_{jb} ~\delta_{ka} \right) \langle \epsilon_a | \mathcal{L}( |\epsilon_c \rangle \langle \epsilon_d |) | \epsilon_b \rangle \nonumber \\ &
+ \left( \delta_{jd}~ \delta_{kc}~ \delta_{la} ~\delta_{mb} + \delta_{ld}~ \delta_{mc} ~\delta_{ja} ~\delta_{kb} \right) \langle \epsilon_b | \mathcal{L}( |\epsilon_d \rangle \langle \epsilon_c |) | \epsilon_a \rangle \nonumber \\ &
- \left( \delta_{jc}~ \delta_{kd}~ \delta_{la} ~\delta_{mb} + \delta_{lc}~ \delta_{md} ~\delta_{ja} ~\delta_{kb} \right) \langle \epsilon_b | \mathcal{L}( |\epsilon_c \rangle \langle \epsilon_d |) | \epsilon_a \rangle \}
\end{align}
\bibliography{quantum}

\end{document}